# Open Multi-Access Network Platform with Dynamic Task Offloading and Intelligent Resource Monitoring

Takuji Tachibana, Kazuki Sawada, Hiroyuki Fujii, Ryo Maruyama, Tomonori Yamada, Masaaki Fujii, and Toshimichi Fukuda

T. Tachibana and K. Sawada are with University of Fukui, Fukui, 910-8507 Japan, e-mail: takuji-t@u-fukui.ac.jp.

H. Fujii, T. Fukuda, R. Maruyama, T. Yamada, and M. Fujii with Fujitsu Limited.

Abstract

We constructed an open multi-access network platform using open-source hardware and software. The open multi-access network platform is characterized by the flexible utilization of network functions, integral management and control of wired and wireless access networks, zero-touch provisioning, intelligent resource monitoring, and dynamic task offloading. We also propose an application-driven dynamic task offloading that utilizes intelligent resource monitoring to ensure effective task processing in edge and cloud servers. For this purpose, we developed a mobile application and server applications for the open multi-access network platform. To investigate the feasibility and availability of our developed platform, we experimentally and analytically evaluated the effectiveness of application-driven dynamic task offloading and intelligent resource monitoring. The experimental results demonstrated that application-driven dynamic task offloading could reduce real-time task response time and traffic over metro and core networks.

I. Introduction

Cutting-edge wireless hardware technologies that satisfy high-frequency bands, ultra-multi-element antennas, and multiplexing have been developed for fifth-generation (5G) mobile communication and beyond 5G (B5G) systems [1]. Optical access network technologies such as next-generation passive optical network 2 (NG-PON2) also improve transmission speeds [2]. Wired and wireless access networks can provide sufficient end-to-end quality of services (QoS) for users [3]. A network system should allow users to utilize various network services as an increasingly important part of social infrastructure without being aware of the access network being wired or wireless [4]. An optimal management and control technology is needed to integrate wired and wireless access networks.

A network system should be designed with network components and devices from various vendors to satisfy the above requirements. It is also necessary to incorporate open-source hardware and software as network components and devices [5,6]. In open network systems, open controllers and orchestrators support end-to-end control, perform software componentization, and ensure interoperability and flexibility. In addition, zero-touch provisioning is essential for reducing the time required for service implementation and effective utilization of access networks [7].

Computing resources also affect the end-to-end QoS of network services. Transmission delays and load must be drastically reduced using cloud and edge computing collaboratively, to process projected large amounts of data for future network services while satisfying end-to-end QoS [8,9].

In this study of network services in the 5G and B5G era, we constructed an open multi-access network platform with the following service features:
- Flexible utilization of network functions
- Integral management and control of wired and wireless access networks
- Zero-touch provisioning
- Dynamic task offloading, with intelligent resource monitoring

Dynamic task offloading and intelligent resource monitoring are implemented for the collaborative processing of tasks in the edge and cloud servers. We propose an application-driven dynamic task offloading that uses intelligent resource monitoring. For performing the application-driven dynamic task offloading, we developed a mobile application, which is an augmented reality (AR) game [12], and an edge-server application for the open multi-access network platform. To investigate the feasibility and availability of our developed platform, we experimentally and analytically evaluated the effectiveness of dynamic task offloading and intelligent resource monitoring.

This study is one of the latest initiatives in constructing a multi-access network platform that utilizes wired and wireless access networks. Especially, the contributions of this study are as follows:
- Constructing a multi-access network platform using open-source hardware and software
- Integrating multi-access network platform and application for task offloading
- Implementing application-driven dynamic task offloading, intelligent resource monitoring, and their cooperation
- Experimental and analytical evaluation of dynamic task offloading in the developed platform

Application-driven dynamic task offloading can be developed and utilized independently of other applications, and the offloading process has been designed specifically for use in an open multi-access network platform, unlike many dynamic task offloading approaches. Hence, it is a promising approach for task offloading in the open multi-access network platform. This article explains how the open multi-access network platform can be developed and investigates the performance of application-driven dynamic task offloading.

## II. Open Multi-Access Network Platform

In this section, we describe the components of the open multi-access network platform for 5G and B5G services. The platform consists of (1) optical access networks that use passive optical networks (PON), (2) wireless access networks that use fourth-generation (4G) long term evolution (LTE) and Wi-Fi technologies, (3) fabric networks that connect the access networks to both the edge clouds and core and metro networks, (4) edge clouds that include edge servers, and (5) management servers that provide managerial and orchestration functions, as well as virtualized network functions (VNFs). The platform is built entirely of open-source hardware and software.

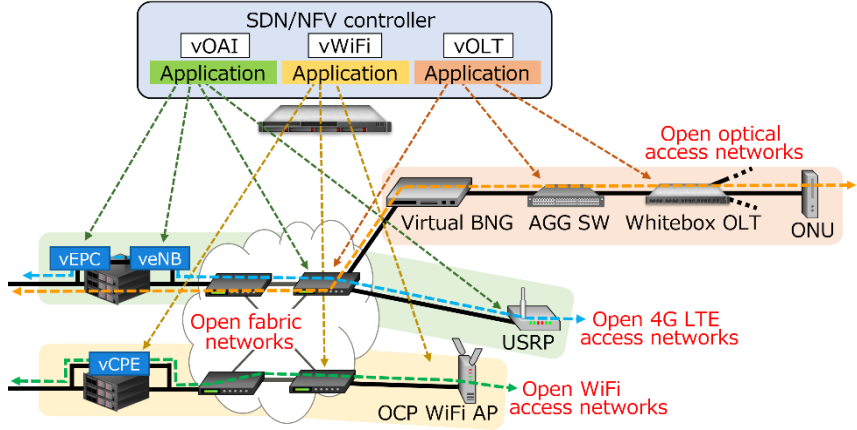

Fig. 1. Control and management of wired and wireless access networks. OLT: optical line terminals, ONU: optical network units, AGG SWs: aggregation switches, BNG: broadband network gateway, USRP: universal software radio peripheral, AP: access point.

### A. Open Optical Access Networks

Open-wired access networks are constructed with 10 Gigabit-capable symmetric (XGS)-PON technology. As shown in Fig. 1, these networks are constructed with white-box optical line terminals (OLTs) and optical network units (ONUs). The optical access networks are connected to the fabric network via aggregation switches (AGG SWs) and a virtual broadband network gateway (BNG).

The optical access networks are controlled and managed by software-defined networking (SDN)-enabled broadband access (SEBA), which includes multiple components such as vOLT hardware abstraction (VOLTHA), open network operating system (ONOS), and network edge mediator (NEM). ONOS is an SDN controller, and SEBA is a variant of residential-central office re-architected as a data center (R-CORD). CORD replaces the traditional central office infrastructure with open software and commoditized hardware building blocks and is a use case for eXtensible cloud Operating System (XOS), a model-based platform for defining, composing, and controlling services. VOLTHA is used to manage white box OLTs for XGS-PON. The management servers in our platform deploy VOLTHA, ONOS, SEBA, and CORD.

### B. Open Wireless Access Networks

Open wireless access networks are deployed based on two architectures types: 4G-LTE and Wi-Fi, as shown in Fig. 1. 4G-LTE consists of evolved universal terrestrial radio access network base stations (eNBs), remote radio heads (RRHs), and evolved packet cores (EPCs). Therefore, open wireless access networks consist of universal software radio peripheral (USRP) that operates as an RRH, virtualized eNBs (veNBs), and virtualized EPC (vEPCs). On our platform, veNBs and vEPCs are deployed as VNFs in the edge servers within the open edge cloud.

The Wi-Fi access networks are constructed from open compute project access points (APs) and virtualized customer premises equipment (vCPEs). The vCPEs, which provide network functions such as domain name system, firewall, and network address translation, are deployed in the edge servers as VNFs.

### C. Open Fabric Networks and Open Edge Clouds

The interconnecting open fabric networks are designed as non-blocking, leaf, and spine topology white box layer 2 switches. This topology was designed using Trellis, which is an open-source software for multi-purpose L2/L3 leaf-spine switching fabric, and each switch is controlled using ONOS. The fabric networks are connected to the open edge clouds, and the open optical and wireless access networks connect to the open edge clouds via fabric networks. The optical and wireless access networks can also connect to the metro and core networks via the fabric networks.

The open edge clouds, including edge servers, are constructed using an open virtual switch (OVS) with a data plane development kit (DPDK). CORD is used as a network function virtualization (NFV) controller with an orchestration function to manage and deploy veNBs, vEPCs, and vCPEs in the edge servers as VNFs. Other VNFs and applications for new network services can also be deployed in edge servers.

### D. Open Management and Orchestration

In the management server, CORD manages and controls the open optical and wireless access networks, open fabric networks, and open edge clouds in an integrated manner. Some automated management operations are introduced as zero-touch provisioning to quickly provide advanced and diversified network services.

A logical path is created automatically for each access network, and flexible routing and switching control for automatic generation and configuration of service tenants are automatically realized. For open optical access networks, first, an ONU is registered in CORD to create a logical path from the ONU. By connecting the ONU to a whitebox OLT in CORD, a PON port is assigned to the whitebox OLT, and the logical path is created between the ONU and the whitebox OLT. For open 4G-LTE access networks, a logical path between a vEPC and a USRP is created by executing a dedicated shell script in a veNB. Specifically, a path is created from the veNB to both the vEPC and USRP, and the logical path is created from the two paths. For open WiFi access networks, a logical path is created between an Open WiFi AP and a vCPE by processing data forwarding from the Open WiFi AP to the vCPE. This process can also be performed by executing a dedicated shell script.

Closed-loop management is also implemented for optimally controlling the resources in the open multi-access network platform. For example, the route of each flow is controlled between two switches by ONOS, and the container migration is performed by CORD. The closed-loop management utilizes the information on transmission delay, CPU utilization rate, etc., which are measured and collected by our intelligent resource monitoring, which is explained in Section III-B.

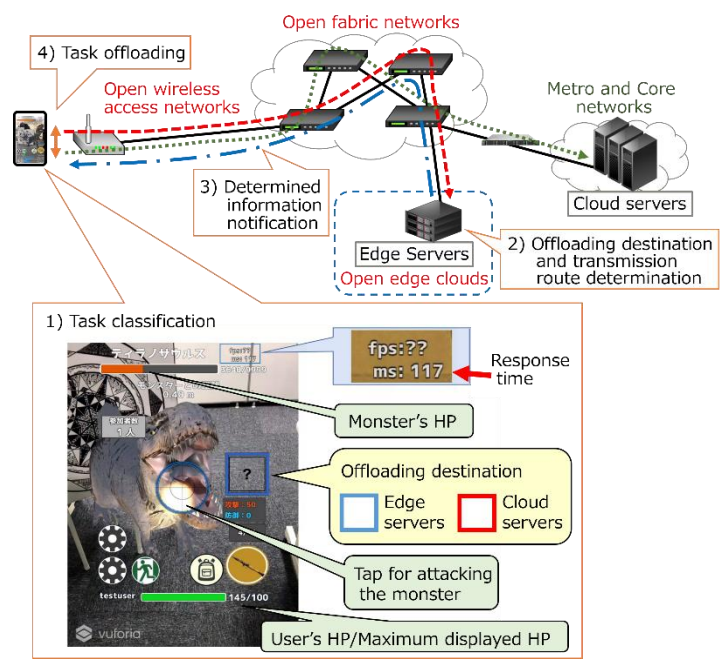

Fig. 2. Processing steps of dynamic task offloading.

## III. Application-driven Dynamic Task Offloading and Intelligent Resource Monitoring

In our multi-access network platform, edge servers within the edge clouds can effectively process data within a few milliseconds for 5G and B5G services. The processing in edge servers can also reduce the amount of traffic over metro and core networks. Edge servers have limited computing capabilities in comparison to cloud servers; hence, they may fail to process some computation-intensive tasks with a certain time constraint. Therefore, it is important to utilize edge and cloud servers collaboratively. The dynamic task offloading and intelligent resource monitoring on our platform serves this purpose.

### A. Processing Steps of Application-Driven Dynamic Task Offloading

Application-driven dynamic task offloading is performed by integrating the open multi-access network platform with applications as follows:
- Step 1: Task classification
- Step 2: Offloading destination and transmission route determination
- Step 3: Determined information notification
- Step 4: Task offloading

Fig. 2 illustrates how these four steps are performed. Unlike many dynamic task offloading approaches, these steps are specifically designed for use in an open multi-access network platform. The mobile application and server applications handle the majority of the processes in this algorithm, with the exception of intelligent resource monitoring. The application provider can determine and implement the details of each process independently of other applications.

In Step 1, tasks are classified into three categories: firm real-time, soft real-time, and non-real-time tasks [14]. The non-real-time tasks are processed on a best-effort basis without time-bound deadlines, and the processing result is always correct. Soft real-time tasks have time bounds, but process results can be used as valid results even after time-bound deadlines. Soft real-time tasks are prioritized over non-real-time tasks and processed quickly. On our platform, firm real-time tasks are always processed in the edge servers to receive process results within time-bound deadlines. Non-real-time tasks are always processed in the cloud servers. Given a soft real-time task, the priority is to process the task on the edge servers as soon as possible. If the non-real-time tasks are to be processed on idle edge servers, the tasks are classified as soft real-time tasks. When the capacity of the edge servers is not sufficient, the QoS of the application is significantly degraded because the response time of firm real-time tasks is significantly increased. In application-driven dynamic task offloading, the task classification is determined and performed in the mobile application or the edge server application independently of other applications.

In Step 2, the edge servers determine the offloading destination and transmission route for processing soft real-time tasks. This process is performed with intelligent resource monitoring, explained in Section III-B, based on the status of the multi-access network platform. To appropriately determine the offloading destination and transmission route, some optimization problems have been formulated, and algorithms for solving these problems have been proposed [15]. For the application-based dynamic task offloading, distributed approaches can be utilized by extending server applications such that multiple edge servers can collaborate.

In Step 3, edge servers notify each application about the determined information for each task. The determined information is added to a packet of task processing results or is sent using a dedicated packet. When the overhead of destination notification is large, the mobile application may send the notification to a limited number of mobile terminals. After an application receives the information on the offloading destination and transmission route, the application sets the destination and transmission route for a soft real-time task in Step 4. The application then sends the task to the multi-access network platform. The destination of only soft real-time tasks is changed; therefore, task classification affects the performance of the proposed algorithm. It is important to determine and perform task classification adequately. Furthermore, by extending the algorithm, an edge server can be chosen from among multiple edge servers. The task of offloading from one edge server to another is carried out.

### B. Intelligent Resource Monitoring with INT and Server Monitoring

The intelligent resource monitoring in Step 2 utilizes in-band network telemetry (INT) [10] and server monitoring [11] for edge servers and access/fabric networks, as shown in Fig. 3. In this case, INT is implemented using programming protocol-independent packet processors (P4) and server monitoring with Prometheus and Grafana.

INT collects timestamps at each P4 switch and smart network interface card (NIC) and stores the information in a new timestamp field defined for each packet. In Fig. 3, Timestamp 1, Timestamp 2, and Timestamp 3 are stored in a packet of each task at P4 Switch 1 (SW 1), P4 Switch 2 (SW 2), and P4 Switch 3 (SW 3), respectively. P4 switches also collect timestamps for a packet of task processing results, which are transmitted from the edge or cloud servers.

When the offloading destination of a packet is the edge servers, the packet is copied at SW 2 adjacent to the edge servers, and the copied packet is sent from SW 2 to the closest telemetry server to the edge servers, which, for illustration, is labeled Telemetry Server 1. The timestamp fields for INT are deleted from the packet at SW 2, and the original packet is sent to the edge servers for task offloading. If the offloading destination is a cloud server, SW 3 and Telemetry Server 2 collect the timestamps and delete the timestamp field.

In addition, each packet has a unique ID number that identifies each task and the corresponding processing results. For each task, Telemetry server 1 derives the response time including the task processing time and transmission delays from the difference between Timestamp 1 of the task and Timestamp 1 of the task processing results. This response time is sent to the edge servers and is compared with the threshold in the application-driven dynamic task offloading. In Fig. 4, the top graph shows the number of packets for each transmission delay between two P4 switches. The transmission delay of each packet was derived by Telemetry servers 1 and 2. As expected, the transmission delay between SW 1 and SW 2 for the edge servers is smaller than that between SW 1 and SW 3 for the cloud servers.

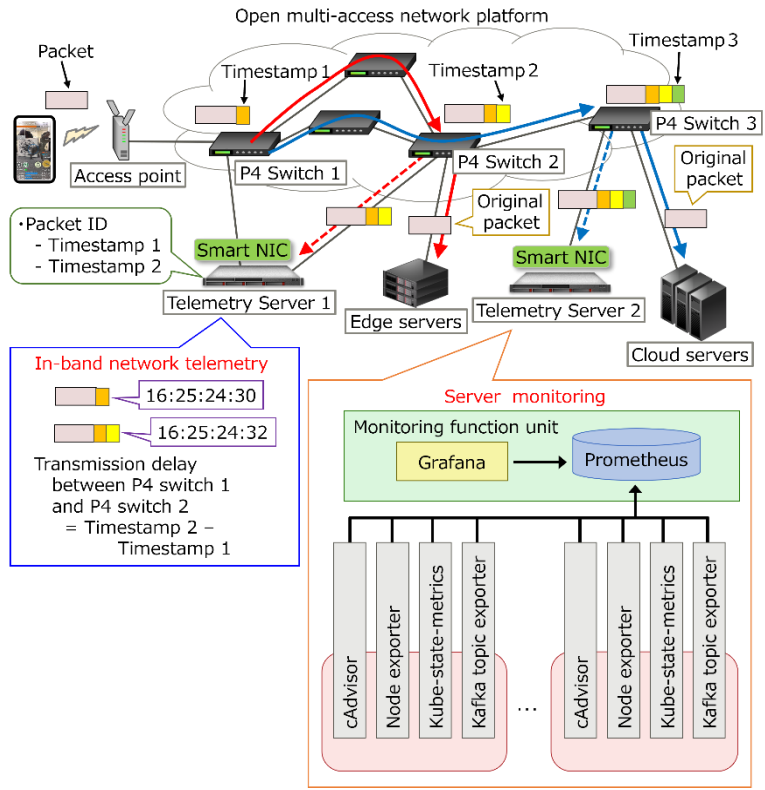

Fig. 3. Intelligent resource monitoring with in-band network telemetry and server monitoring.

Each edge server has cAdvisor, node exporter, Kube-state-metrics, and Kafka topic exporter for server monitoring, and the monitoring function unit uses Prometheus and Grafana. A cAdvisor collects information about resource usage and performance characteristics of running containers, and a node exporter measures machine resources such as CPU, memory, and hard disk. A Kube-state-metrics is an open-source utility used to monitor the state of a couple of Kubernetes objects in each Kubernetes cluster. A Kafka topic exporter is used to build real-time streaming data pipelines to send the monitoring information to Prometheus. Grafana is used to visualize the monitoring results obtained with Prometheus. The bottom graph in Fig. 4 shows the CPU utilization rate for a user application and system events. By the resource monitoring, it is clear that the user application utilizes CPU more than the system events in the edge servers.

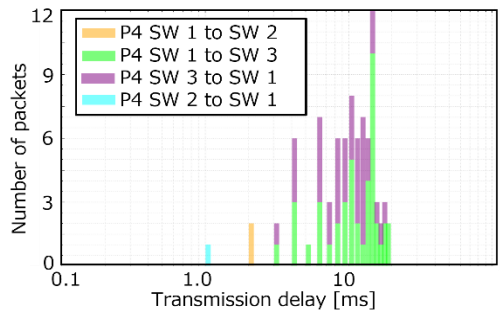

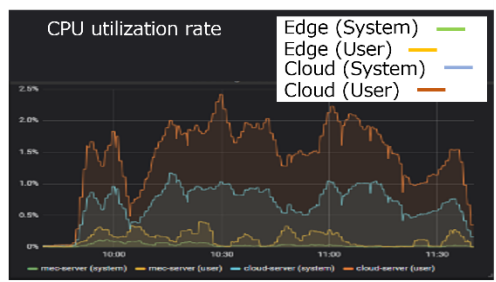

Fig. 4. Examples of Intelligent Resource Monitoring.

## C. Implementation

We developed a multi-mobile application using Unity and the Vuforia library as an application that integrates with the open multi-access network platform for application-driven dynamic task offloading. This application is an AR game in which many users can participate [12].

In this application, when a user starts this application, the application connects to a cloud server, and the user's information is registered in the MySQL database of the cloud server. When the user selects a monster on the screen, the related information is sent to the cloud servers as a non-real-time task. After the application receives the latest information on the monster, the selected monster is displayed on an AR box, and the battle starts for the user. All users share the same information on the monster in this game, and each user attacks the monster by tapping the screen. A successful attack always sends the result to an edge server as a firm real-time task. The edge server manages the information on the monster's HP. This information is updated periodically based on the received results of successful attacks, and the edge servers return the updated monster's information to the users. In this application, users can use items to support the battle. Such items can be processed as a soft real-time task either by the edge or cloud servers.

For the offloading destination and transmission route determination in Step 2, the algorithm was implemented in the edge servers. The edge servers receive the JSON file, including information on the intelligent resource monitoring from telemetry servers. In the current version, the offloading destination of the soft real-time task is edge servers when the response time is equal to or smaller than the threshold to satisfy the time-bound deadlines for the end-to-end QoS. Otherwise, the offloading destination is cloud servers because the edge servers should process the firm real-time tasks. The edge servers add the information on the determined offloading destination and transmission route into the packet of the updated monster's information and send it to the application periodically at Step 3. Finally, the application sends the soft real-time task to the edge or cloud servers based on the determined information notification. Here, both the determined information notification and task offloading are performed with WebSocket in Steps 3 and 4.

## IV. Numerical Examples

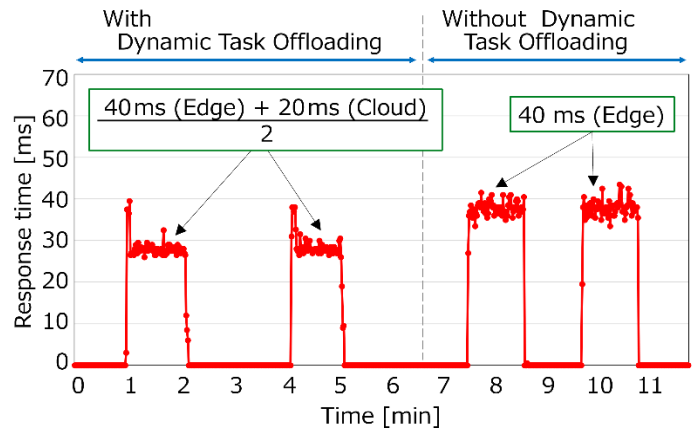

Fig. 5. Impact of application-driven dynamic task offloading on the average response time.

### A. Experimental Evaluation

For the developed open multi-access network platform shown in Fig. 3, we evaluated the performance of application-driven dynamic task offloading and intelligent resource monitoring for the above-mentioned mobile application. In this experiment, the processing delay of a task in the edge and cloud servers can be changed using a traffic control command. The load of the edge servers is also adjustable by changing the number of available CPU cores, resulting in an increase or decrease in the processing delay.

The offloading destination of a firm real-time task is always the edge servers, but the offloading destination of a soft real-time task can be changed if application-driven dynamic task offloading is utilized. The threshold value for application-driven dynamic task offloading was set to 25 ms to satisfy the time-bound deadlines for the end-to-end QoS of the firm and soft real-time task. When the response time is shorter than the threshold, the edge servers can decide the offloading destination for a soft real-time task as edge servers; otherwise, the cloud servers process the soft real-time task.

Fig. 5 shows the average response time, measured at P4 Switch 1 with INT, for the processing result of a soft real-time task. For simplicity, the number of soft real-time tasks for using support items was set as the same as the number of firm real-time tasks, which are the attacks, by modifying the application. In this experiment, application-driven dynamic task offloading was used from 0 s to 6 min 48 s and not used afterward. The processing time of the edge servers varied between approximately 0 ms and 40 ms by changing their load. A processing time of approximately 40 ms was set from 59 s to 2 min 7 s, 4 min 9 s to 5 min 6 s, 7 min 32 s to 8 min 37 s, and 9 min 47 s to 10 min 56 s. The processing time of the cloud servers was approximately 20 ms, which is lesser than that of the edge servers with high loads.

In Fig. 5, the response times of some tasks are approximately 40 ms immediately after 59 s or 4 min 9 s, even if the application-driven dynamic task offloading is used. This is because the tasks are sent to the edge server with high loads before the destination is changed. After the destination is changed by the application-driven dynamic task offloading, the average response time is approximately 30 ms, whereas the response time of firm real-time task offloading and soft real-time task offloading is approximately 40 ms and 20 ms, respectively. When application-driven dynamic task offloading is not available, the average response time is approximately 40 ms because the soft real-time task's response time is also approximately 40 ms. Thus, when compared to cases without application-driven dynamic task offloading, using application-driven dynamic task offloading reduces average response time. Note that the response time for firm real-time tasks is also reduced by the application-driven dynamic task offloading because the edge server loads can be decreased. The effectiveness of the application-driven dynamic task offloading was confirmed even when the number of soft real-time tasks and the number of firm real-time tasks is different.

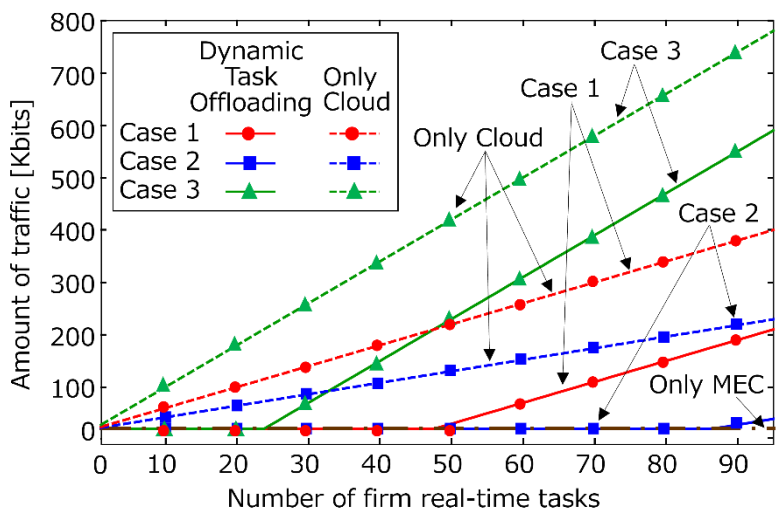

Fig. 6. Impact of application-driven dynamic task offloading on the amount of traffic over the metro and core networks.

## B. Analytical Evaluation

We also analyze the performance of our application-driven dynamic task offloading in terms of the amount of traffic transmitted over the metro and core networks. The firm real-time and soft real-time tasks were transmitted from applications, and synchronized data were transmitted between the edge and cloud servers every 0.5 s. The maximum number of tasks that can be processed per 1.0 s in the edge servers was 100, and the packet size of each task and synchronized data was 10 Kbits. These parameters were set based on the above-mentioned experimental evaluation. We considered three cases in terms of the ratio of the number of soft real-time tasks to the number of firm real-time tasks; the ratio considered were 1.0, 0.5, and 3.0 for cases labeled 1, 2, and 3, respectively. Note that the total number of tasks for case 2 was the smallest and that for case 3 was the largest when the number of firm real-time tasks is the same.

The amount of traffic [Kbits] over the metro and core networks is plotted against the number of firm real-time tasks for 1.0 s in cases with and without application-driven dynamic task offloading in Fig. 6. The graph also depicts the performance of task offloading for only cloud servers. If the task offloading was performed for only the cloud servers, the amount of traffic increased linearly when the number of processed tasks was larger than zero. With application-driven dynamic task offloading, the amount of traffic over the metro and core networks was suppressed when the edge server processes soft real-time tasks. As the ratio of soft real-time tasks increases, the amount of traffic over the metro and core networks increases even when the amount of firm-real time tasks is small, i.e., case 3. However, the number of firm real-time tasks processed in the edge servers does not decrease because soft real-time tasks were processed in the cloud servers preferentially.

Because the cloud servers were not used in the absence of dynamic task offloading, the amount of traffic does not increase regardless of the number of firm real-time tasks. However, the response time for all tasks became excessively long, and the end-to-end QoS could not be met. As a result, dynamic task offloading effectively reduces traffic over metro and core networks while meeting the end-to-end QoS of firm real-time tasks.

## V. Conclusion and Future Direction

In this study, we developed an open multi-access network platform with open-source hardware and software. Moreover, we proposed and implemented a dynamic offloading framework and intelligent resource monitoring. We developed a mobile application and experimentally investigated the performance of application-driven dynamic task offloading for our developed mobile application in the open multi-access network. We then analyzed the effectiveness of application-driven dynamic task offloading in three cases. Numerical results demonstrated that application-driven dynamic task offloading could reduce the response time of soft real-time tasks while also reducing traffic over metro and core networks.

We are currently expanding the open multi-access network platform. Open devices and software usable in pluggable OLTs and NG-PON2 will be considered for wired access networks. In terms of wireless access networks, we will use open devices and software for Wi-Fi 6, which is IEEE 802.11ax, and open devices and software for 5G wireless technology. We are also implementing a more sophisticated heuristic algorithm for application-driven dynamic task offloading and extending our mobile application. To improve reliability, open management and orchestration will be expanded to consider the redundancy of various server and network resources.

In the future, programmable switches that support more accurate time synchronization and hardware acceleration technology for the virtualization infrastructure will be introduced for open fabric networks. We also plan to introduce an open technology for the flexible control of network and computer resources for network virtualization. Furthermore, open technologies for optimal deployment and resource allocation of VNFs in management and edge servers will be considered, and the orchestrators and controllers will be extended for zero-touch provisioning operations.

## Acknowledgment

The research was funded by the Commissioned Research of the National Institute of Information and Communications Technology (NICT), JAPAN, as part of the ``Open Multi-access Network Platform for 5G and Beyond 5G Services.'' We also thank Mr. Naoya Okamoto for his assistance in developing our mobile application.

platform," in Proc. 2019 IEEE Intl. Conf. on Consumer Electronics-Taiwan, May 2019.

**Takuji Tachibana** received the Dr. Eng. from Nara Institute of Science and Technology, Japan, in 2004. Since 2019, he has served as a professor at University of Fukui, Japan. His research interests include network virtualization, optical networking, and performance analysis of communication systems.

**Kazuki Sawada** received his M. Eng. from University of Fukui, Japan, in 2021.

**Hiroyuki Fujii** joined with Fujitsu Ltd., Kawasaki, Japan, in 2002 where he has been engaged in the research and development of optical communication systems, 3G / 5G base stations, SDN / NFV systems.

**Ryo Maruyama** joined with Fujitsu Limited, Kawasaki, Japan, in 1999 where he engaged in the research and development of optical communication systems. Since 2017, he has been engaged in business planning of mobile network systems.

**Tomonori Yamada** joined with Fujitsu Access, Kawasaki, Japan, in 2005. Since 2017, he has been with Fujitsu Limited, Kawasaki, Japan, where he has been engaged in the research and development, and business planning of optical communication systems.

**Masaaki Fujii** joined Fujitsu Limited, Kawasaki, Japan, in 2009. He has been engaged in the research and development of optical transport systems and network virtualization.

**Toshimichi Fukuda** joined Fujitsu Limited, Kawasaki, Japan, in 1992. He has been engaged in software development for optical network systems, data center network, and virtualized network systems.